\documentclass[10pt, conference, compsocconf]{IEEEtran}
\usepackage{cite}

\ifCLASSINFOpdf
  \usepackage[pdftex]{graphicx}
\else
\fi

\usepackage{graphicx}
\usepackage{booktabs} 
\usepackage{fancyhdr}

\usepackage{amsthm}
\usepackage{amsmath}
\usepackage{caption}
\usepackage[ruled,linesnumbered]{algorithm2e}
\usepackage{algpseudocode}
\usepackage{epstopdf}
\usepackage{subfigure}

\begin{document}
\title{An Efficient Graph Accelerator with Parallel\\ Data  Conflict Management}

\author{\IEEEauthorblockN{Pengcheng Yao}
\IEEEauthorblockA{Huazhong University of Science and Technology\\
pcyao@hust.edu.cn}
}

\maketitle

\begin{abstract}

Graph-specific computing with the support of dedicated accelerator has greatly boosted the graph processing in both efficiency and energy. Nevertheless, their data conflict management is still sequential in essential when some vertex needs a large number of conflicting updates at the same time, leading to prohibitive performance degradation. This is particularly true for processing natural graphs.


In this paper, we have the insight that the atomic operations for the vertex updating of many graph algorithms (e.g., BFS, PageRank and WCC) are typically incremental and simplex. This hence allows us to parallelize the conflicting vertex updates in an accumulative manner. We architect a novel graph-specific accelerator that can simultaneously process atomic vertex updates for massive parallelism on the conflicting data access while ensuring the correctness. 
A parallel accumulator is designed to remove the serialization in atomic protection for conflicting vertex updates through merging their results in parallel. 
 Our implementation on Xilinx Virtex UltraScale+ XCVU9P with a wide variety of typical graph algorithms shows that our accelerator achieves an average throughput by 2.36 GTEPS as well as up to 3.14x performance speedup in comparison with state-of-the-art ForeGraph (with single-chip version). 
\end{abstract}

\IEEEpeerreviewmaketitle

\section{Introduction}
Graph processing plays an important role in many real-world applications, e.g., ranking the web sites~\cite{shun2013ligra}, analysing the social networks~\cite{teixeira2015arabesque}, and streaming applications~\cite{liao}. Therefore, a large number of research efforts have been made to build the dedicated hardware that can execute graph applications with more efficiency than what the general-purpose processors and systems can provide~\cite{nurvitadhi2014graphgen, ham2016graphicionado, ozdal2016energy, dai2017foregraph}.

Despite these efforts, the graph algorithms may still suffer from a considerable performance impact caused by the atomic protections. During the graph iteration, each vertex sends its value to all associated vertices. Therefore, it is common that many vertices may read/write the same vertex simultaneously, needing a significant number of atomic protections in existing graph accelerators for preserving the correctness. This performance overhead arising from the atomic operations can be as much as nearly half of total graph execution, as demonstrated in previous work~\cite{nai2017graphpim, wu2015g} and also witnessed in our motivating study in Section 2. 

Much effort has been put into reducing the atomic overhead. By offloading the atomic operations to specialized memory (e.g., hybrid memory cubes~\cite{lee2015bssync,nai2017graphpim}), data access overhead can be reduced. Speculative lock elision can expose the fine-grained parallelism due to inappropriate atomic protection~\cite{Herlihy1993Transactional}. Recent studies also attempt to reduce the number of atomic operations by a series of sophisticated preprocessing, e.g., graph partition~\cite{dai2017foregraph} and dynamic scheduling~\cite{ozdal2016energy}. Unlike these previous work that concentrates on optimizing the individual atomic overhead, this work focuses on the totally-sequential performance impact between atomic operations, which is under-studied in graph processing.


Interestingly, graph processing for many graph algorithms (e.g., BFS, PageRank and WCC) shows a significant, common feature for their atomic operations: 1) {\em incremental}--the atomic operations follow the commutative and associative law, 2) {\em simplex}--all atomic operations are similar. Instead of enforcing sequential execution of conflicting operations as traditional designs, this unique observation in graph processing enables to parallelize massive conflicting vertex updates in an accumulative manner in the sense of simultaneously processing multiple operations and merging the results in parallel. In this paper, we are addressing how we can design such an efficient accumulator for parallelizing the conflicting data accesses for vertex updating in graph processing. 

We propose a novel accelerator that can simultaneously process multiple atomic operations for parallelizing the vertex updates with a data conflict while ensuring the correctness. Considering that the real-world graphs generally follow the power-law distribution~\cite{gonzalez2012powergraph}, a specialized accumulator is designed to distinguish the processing of low-degree and high-degree vertices. Internally, it executes multiple low-degree vertices in parallel for efficient edge-level parallelism and limits the vertex parallelism for the high-degree vertices to avoid frequent synchronization. To keep the architecture balanced, our accelerator is built with a high-throughput on-chip memory to provide efficient vertex access for the accumulator. The memory evenly distributes the requests based on a rearranging mechanism and process them in an out-of-order manner to ensure an efficient throughput.


The contributions of this work are summarized as follows:


\begin{itemize}
\item We study a wide range of graph workloads and perform a detailed analysis on their atomic operations. We demonstrate that their distinct characteristics enable the parallel execution for conflicting vertex updates.
\item We propose a graph-specific accelerator which supports parallel execution of atomic operations. A parallel accumulator is designed to guarantee efficient process of vertices with different degrees. A high-throughput on-chip memory is also provided for the efficient use. 
\item We compare our accelerator with the state-of-art ForeGraph. Experimental results with three graph algorithms on six real-world graphs show that our accelerator provides 2.36 GTEPS on average, outperforming ForeGraph by up to 3.14x speedup. 
\end{itemize}

The rest of this paper is organized as follows. In Section 2, we introduce the background of graph processing and provide our motivations and challenges in detail. Section 3 and Section 4 propose our parallel accumulator designs and optimizations in memory subsystem. The evaluation results are presented in Section 5. We survey related work in Section 6 and conclude the paper in Section 7.

\section{Background and Motivation}
This section first reviews the vertex updating mechanism of existing graph accelerators for the conflicting data accesses. We next discuss its potential deficiency for graph processing through a motivating study, finally presenting our approach. 
\subsection{Modern Graph Accelerator and Its Data Conflict Management}


\begin{figure}
\centering
\subfigure[Pseudocode of BFS]{

\begin{minipage}[b]{0.5\textwidth}
\centering
\includegraphics[width=2.7in, height=1.6in]{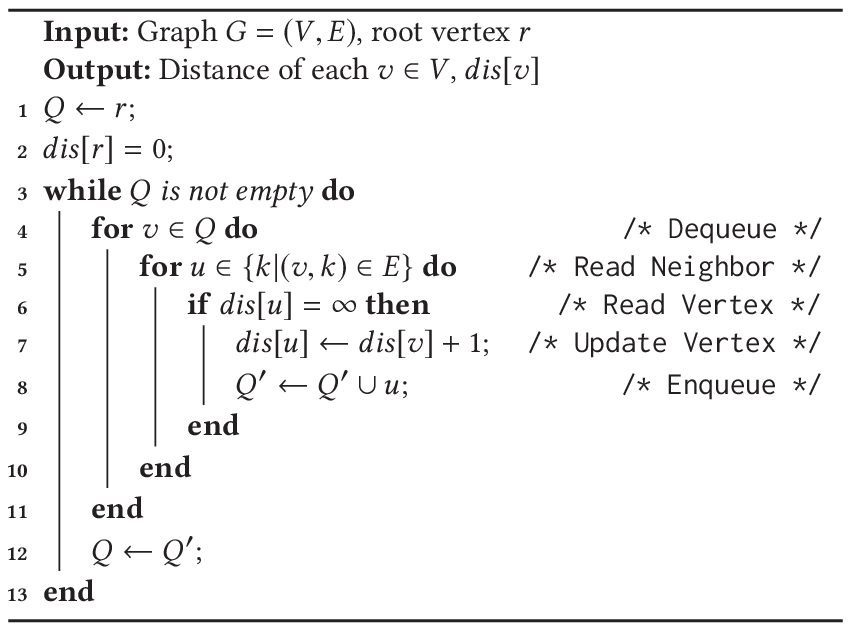}
\label{fig_bfs_code}
\end{minipage}
}

\subfigure[Execution flow of BFS]{
\begin{minipage}[b]{0.5\textwidth}
\centering
\includegraphics[width=2.7in, height=0.7in]{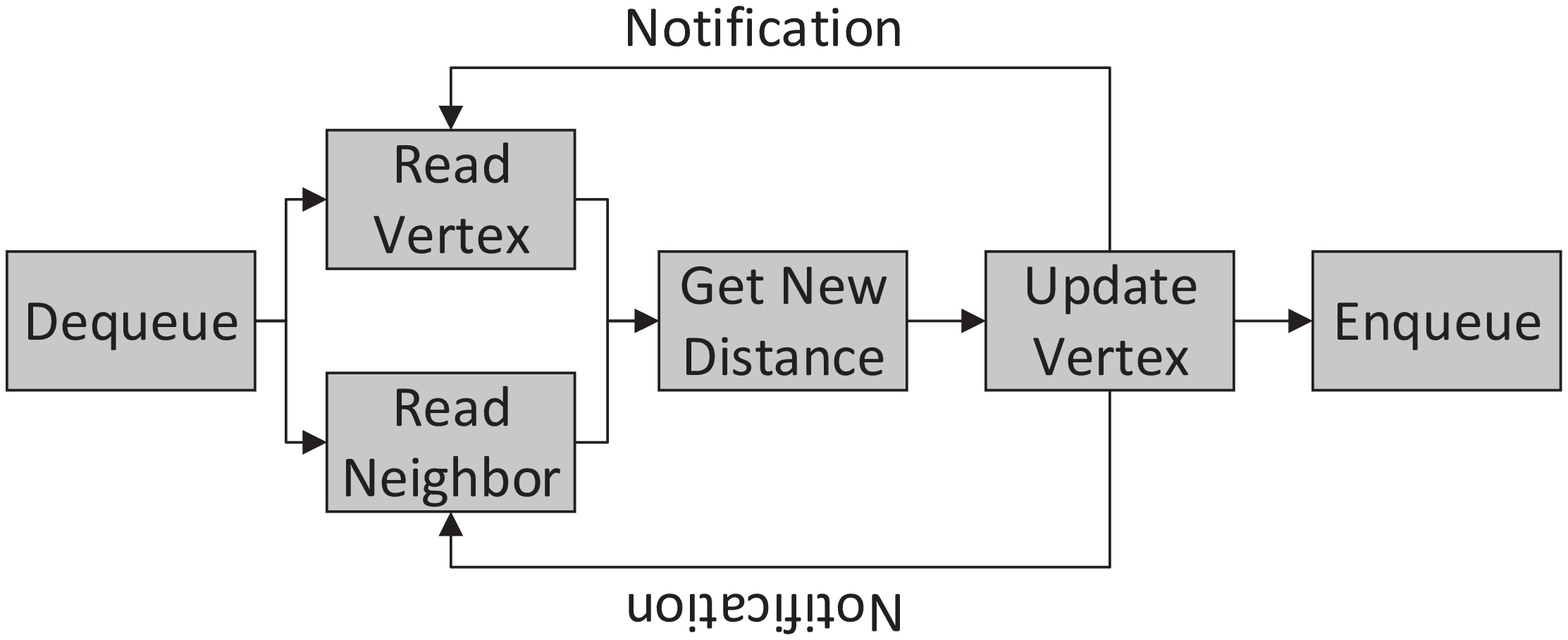}
\label{fig_bfs_atomic}
\end{minipage}
}
\caption{BFS pseudocode and its execution flow} 
\vspace{-1.5em}
\end{figure}


Graph accelerator is a customized hardware that is specially designed for iterating the computation on graphs. In graph representation, each entity is traditionally defined as {\em vertex}, and its connection is defined as {\em edge}. The {\em degree} of a vertex denotes the number of connections it has. The degree distribution is the probability distribution of all degrees. 

In existing graph accelerators with shared memory architecture, all vertices in the graph are shared and also able to be accessed by multiple pipelines. As a result, there is a high coverage of data contentions for graph processing, particularly those vertices associated with a large number of edges. For ensuring the correctness of vertex updating, existing researches often seek to use atomic structures (e.g., content addressable memory~\cite{ham2016graphicionado, ozdal2016energy, pagiamtzis2006content}), which tend to atomically protect the update of each vertex if a conflicting data access to this vertex has been detected at runtime. 

A typical procedure of data conflict management used in many graph accelerators~\cite{ham2016graphicionado, ozdal2016energy} is as follows. Multiple edges of the given vertices will be fetched and sent to the accelerator in each cycle. When receiving these edges, the accelerator will check the pipeline states at first. If an edge is connected with a vertex which is executing in the pipeline, its process would be stalled until the prior one finishes execution. In this way, the same vertex cannot appear in more than one of the pipeline stages for vertex execution at the same time, thus ensuring atomicity.



\subsection{Inefficiency in Graph Processing}
Graph often exhibits the complex connections where any vertex may be shared among different vertices. This is particularly true and serious for nature graph that follows the power-law degree distribution, where most vertices have low degree while a few have extremely large degree~\cite{gonzalez2012powergraph}. Thus, there may involve a high risk that a large number of low-degree vertices simultaneously access the same high-degree vertex, leading to serious data contention. Unfortunately, modern graph accelerators (e.g., ForeGraph~\cite{dai2017foregraph} and Graphicionado~\cite{ham2016graphicionado}) fall short in handling these highly-frequent data conflicts in graph processing due to its serial semantics with atomic protection for vertex updates. 

{\bf Atomic Protection Analysis}\quad 
Figure~\ref{fig_bfs_code} illustrates the pseudo-code of {\em Breadth-First-Search} (BFS). It starts from a root vertex $r$ and iteratively traverses the graph to calculate the shortest distance from the root vertex to other vertices. During the traversal, each vertex $v$ in the scheduling list will receive values $dis[u]$ from its neighboring vertices and update its own data based on these values (Line 7). In the end of the traversal, a new vector $Q^\prime$ is generated and used as the scheduling list of the next iteration.

Because of the atomic protection, these received data from neighboring vertices has to be updated one-by-one in each cycle for preserving the correctness of final result. Figure~\ref{fig_bfs_atomic} shows the execution flow of BFS with atomic protection. Each scheduled vertex will access data from itself and one of its neighbors, and write back the updated data after finishing processing. The data of other neighbors is cached and will not be released to the pipeline before receiving the completion of prior process. In other word, the process inside each vertex is enforced to be sequential for reducing data contention at the cost of performance. 

\begin{figure}
\centering
\includegraphics[width=2.7in, height=1.6in]{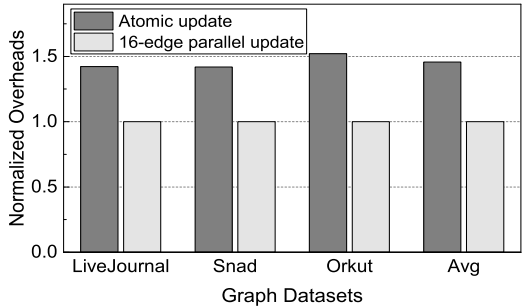}
\caption{Normalized performance overhead caused by sequential atomic operations}
\vspace{-1em}
\label{memory_syn}
\end{figure}

{\bf\em Experimental Demonstration}\quad We further make a set of experiments to investigate how much performance impact may be incurred by atomic protection in graph processing. We use a cycle-accurate simulation to perform the vertex iteration with a parallel update for a maximal set of 16 edges\footnote{The simulation is conducted with a pipelined architecture that is similar to ForeGraph~\cite{dai2017foregraph}. While data width of edges is usually 32-bits in BFS, we set 16-edge parallelism according to the memory access granularity (512-bits). Edge shuffling optimization~\cite{dai2017foregraph} is not covered in our simulation.}. Figure~\ref{memory_syn} depicts the comparative results. 
It is observed that the pure atomic protection leads to a significant performance degradation for all real-world graphs, with 45\% extra memory overheads on average in contrast to 16-edge parallel vertex update. This is particularly true and serious for those graphs that have the greater average degree (e.g., {\em Orkut}). 

{\bf Remark} There are also a number of potential solutions that can be used for reducing the performance impact arising from atomic operations. ForeGraph~\cite{dai2017foregraph} proposes a shuffling mechanism to rearrange the edges with potential data conflicts. ~\cite{ozdal2016energy} excessively schedules destination vertices and sends part of them to the processing unit based on a credit based mechanism. Similarly, the basic idea of the above mechanism is to avoid simultaneously scheduling edges with the same destination vertex. While they can reduce the pipeline stalls caused by atomic protection, they still have sequential process of different edges for the same destination vertex. 

Some work~\cite{ahn2016scalable, nai2017graphpim} uses novel {\em processing-in-memory} (PIM) technology~\cite{gokhale1995processing} to offload the atomic operations to specialized memory region, which reduces the processing time of atomic operations. However, it needs to incorporate with specialized memory architecture and also increases the memory requests since all atomic operations needs to be sent to the memory.

\begin{table}[htbp]
\centering
\caption{Atomic operation types for the vertex update in different graph algorithms} 
\vspace{-1em}
\label{atomic_type}
\begin{tabular}{|c|c|}
\multicolumn{1}{c}{}&\multicolumn{1}{c}{} \\ \hline
{\bf Algorithm}         & {\bf Operation Type} \\ \hline
Breadth-First Search    & CAS if less \\ \hline
Weakly Connected Components    & CAS if less	\\ \hline
Shortest Path           & CAS if less   \\ \hline
PageRank                & Atomic add   \\ \hline
Triangle Counting       & Atomic add  	\\ \hline
Degree Centrality       & Atomic add   \\ \hline
Collaborative Filtering & Atomic add   \\ \hline
\end{tabular}
\vspace{-0.5em}
\end{table}

\subsection{Potential of Accumulator}
The key insight of this work is that atomic operations for many graph algorithms can be parallelized in an accumulative manner. 
Table~\ref{atomic_type} illustrates the typical operations that need an atomic protection for seven popular graph algorithms. We can observe that these atomic operations as a whole have two aspects of significant properties. 

{\bf\em Observation 1}: {\em The atomic operations on different edges follow the commutative and associative law}.

The commutative law means that the execution sequence of the operations has no effect on the result. Associativity ensures the correctness of merging multiple operations. That is, any of the operations can be simultaneously merged without changing the final result. For example, {\em PageRank} follows the atomic-add operations. It updates every vertex by following  $Rank(v) = \varepsilon + \sum_{u \in neighbor(v)} Rank(u) / |neighbor(u)|$, where $\varepsilon$ is a constant. Actually, no matter how we change the sequence of these atomic operations or merge successive atomic operations, the final result can be still consistent.

{\bf\em Observation 2}: {\em The atomic operations for updating the value of conflicting vertex are simple and used repeatedly}.

Taking {\em PageRank} as the example, we find that all of its atomic operations use the same atomic-add to sum up their values to the final result. This similarity allows to use a unique structure to merge all atomic operations.

\begin{figure}
\centering
\includegraphics[width=2.7in, height=1.6in]{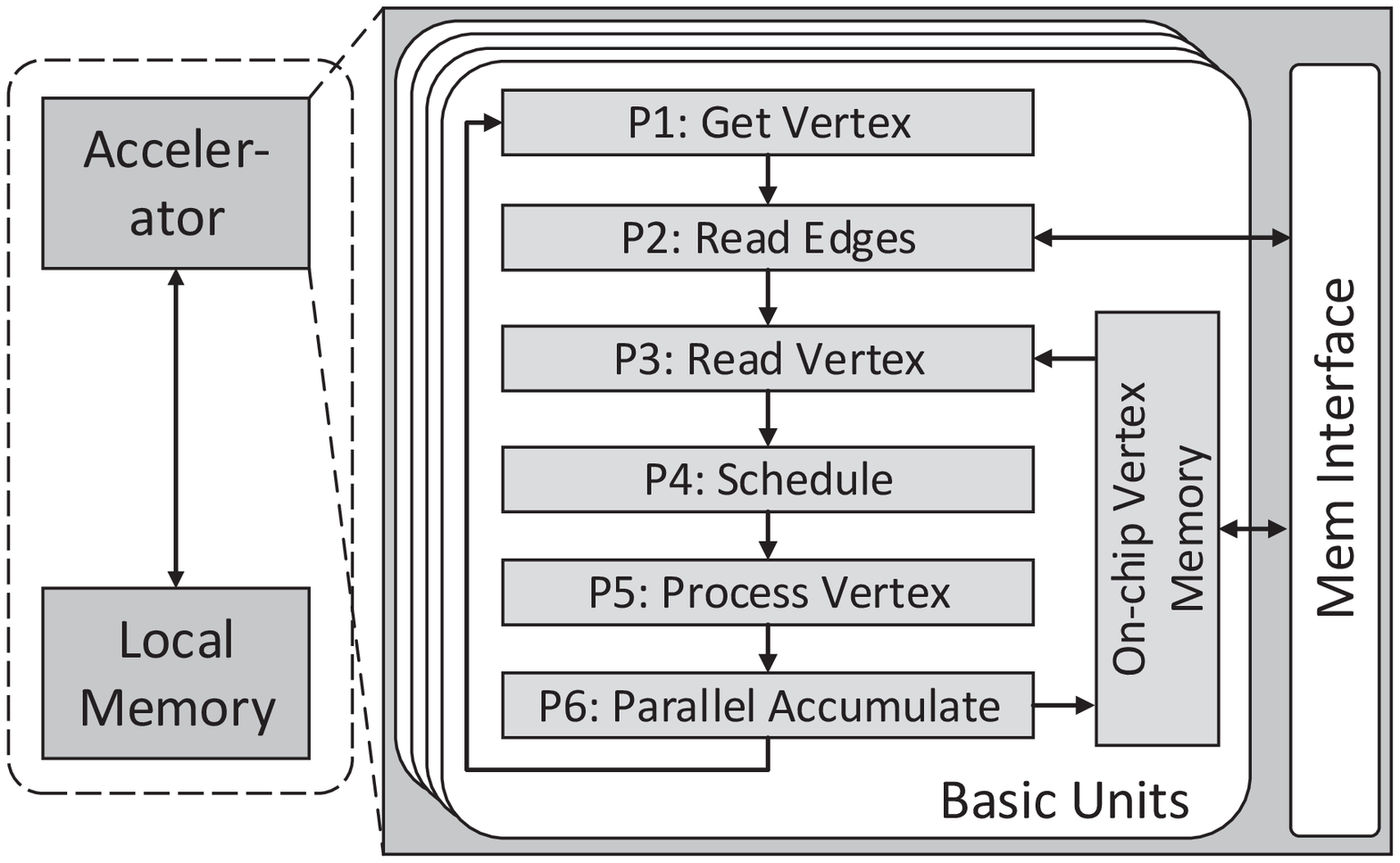}
\caption{Architecture of Graph Accelerator. $P_i$ denotes the $i$th pipeline stage}
\label{architecture_overview}
\vspace{-1.5em}
\end{figure}


These two observations consequently enable us to leverage existing well-developed accumulator to parallelize the vertex update conflicts. Accumulator is a hardware component that merges the inputs into a set of results with specific function. Nevertheless, designing such accumulator for large-scale graph processing remains tremendously challenging.

First, the real-world graph topology is often sparse with a low averaged degree. Although traditional accumulator designs~\cite{knowles2001family, ladner1980parallel, blelloch1989scans} can provide desirable throughput, they often establish a fixed mapping relationship between the inputs and the results. The reality is that the degree of vertices is dynamically changing during the iteration. The accumulator may get incorrect results when simultaneously processing multiple vertices. Therefore, the traditional accumulator can only accumulate the atomic operations of the single low-degree vertex at the same time, leading to extremely low parallelism for graph processing. There remains a significant gap in applying the accumulation ideology into graph processing without losing a wealth of edge-level parallelism. 


Second, natural graphs often follow a power-law distribution. When processing the low-degree vertex, the accumulator is expected to simultaneously process multiple vertices. However, for the high-degree vertices with a large number of edges that can be easily more than millions (e.g., {\tt twitter}), an accumulator with limited width is extremely difficult to handle so many edges simultaneously. If multiple vertices are simultaneously processed in this case, the accumulator will be invoked several times at the cost of increased synchronization overheads. Moreover, it may lead to massive random edge accesses since the edges of these vertices are more likely to be non-sequential. Therefore, there still lacks an effective technique that can improve the synchronization overheads and random accesses for an efficient accumulation.


Third, it is also extremely difficult to predict the non-sequential neighboring vertices of each vertex in real-world graphs. A large number of random accesses have be incurred before invoking the accumulator. Although the accumulator can largely reduce the atomic overheads and provide desirable execution performance, the vertex access remains to be a potential bottleneck and significantly limits the throughput.

\subsection{Architectural Overview}

Figure~\ref{architecture_overview} shows an overview of our accelerator, which is designed in pipeline with six stages in total. These stages basically serve as two major objectives as follows:

{\bf How to Design an Efficient Accumulator} (Section 3): As explained in the challenge discussions, the accumulator generally suffers from the sparse topology and power-law degree distribution in real-world graphs. To achieve desirable performance, the accumulator is expected to efficiently process both of the low-degree and high-degree vertices. 

When processing the low-degree vertex, the accumulator is expected to simultaneously process multiple vertices for efficient parallelism. Since the vertex degrees are mutable during the process, the accumulator should establish a dynamic relationship between the input vertices and the final results to ensure the correctness. 

When processing the high-degree vertex, the number of vertices scheduled should be decreased to avoid random access. Therefore, the accumulator should be dynamically aware of the changes in degree and distinguish the process of different vertices. Furthermore, there is a significant synchronization overhead between the multiple accumulations of the same high-degree vertex, which requires an efficient synchronization mechanism.


{\bf How to Use Accumulator Efficiently} (Section 4): While the accumulator could provide high execution efficiency, the on-chip memory is likely to be a potential performance bottleneck. To keep with the throughput of accumulator, the on-chip memory is required to be partitioned into independent parts to process multiple accesses. Furthermore, considering the randomness in vertex access, the address values of vertices may follow an unbalanced distribution. Consequently, multiple requests will be sent to the same memory part in each cycle, leading to significant throughput degradation. To ensure a high throughput, a specialized mechanism is required to dynamically balance the memory requests for on-chip memory.

\section{Parallel Accumulator Design}
This section discusses the design guideline for a parallel accumulation as well as its core components for the efficiency.
\subsection{Design Philosophy}
Since accumulator is bounded with fixed width, it generally needs to consider two situations where skewed graph vertices with different degrees that can be greater or less than accumulator width, involving different parallel designs.
\subsubsection{\bf Accumulation Design for Low-Degree Vertex} 
As is known, most of vertices for a natural graph have a very few degree which can be often no more than the fixed number of ports for a typical accumulator. It is clear of a necessity to simultaneously process the update values of multiple low-degree vertices at a time for high parallelism.

{\bf Problem Definition:} Assuming $N$ update values, belonging to $M$ vertices, need to be processed at once. This problem can be described by $p_j =  \sum_{1 \le i \le N} a_i \cdot b_{ij}, 1 \le j \le M$, where $p_j$ denotes the accumulated result of vertex $j$. $a_i$ denotes the update value $i$, and $b_j^i$ denotes whether $a_i$ belongs to vertex $j$. The objective is to get $p$ with minimal latency. 

Considering the locality of graph traversal, this problem can be further simplified. During traversal, edges of the same destination vertex are sequentially accessed in common graph representations, e.g., CSR/CSC~\cite{shun2013ligra}. It ensures that update values of the same destination vertex are sequentially received by the accumulator. Therefore, assuming that $C_j = [c_j^1, c_j^2]$ denotes the interval of vertex $j$'s update values in all $a_i$, the function of accumulator could be simplified by $p_j = f(c_j^2)$, where
\begin{align}
\label{compressed_dp}
f(i) = \left \{
\begin{aligned}
f(i - 1) + a_i, & \quad i \notin \{c_1^1, c_2^1, \ldots, c_M^1\} \\
a_i,       & \quad i \in \{c_1^1, c_2^1, \ldots, c_M^1\}
\end{aligned}
\right.
\end{align}

{\bf Solution Discussion:} A naive method for solving this problem is to use a Multi-N-Way~\cite{ma2017garaph} accumulator, which reserves a N-Way accumulator with the binary tree architecture for each vertex. However, its hardware overhead is unacceptable for graph applications. First, its fanouts are too large to implement, which can be up to 8192 when processing a cacheline-width data for 16 vertices. Second, its resource utilization is extremely low since only $N$ among $N \times M$ received values are useful for the real accumulation. 

In Equation~(\ref{compressed_dp}), we find that $f(i) = f(i - 1) + a_i$ is a typical prefix-sum problem, which has been extensively studied in previous work~\cite{sklansky1960conditional, kogge1973parallel, ladner1980parallel, brent1982regular, knowles2001family}. Beyond the prefix-sum problem, a significant problem is that we still need to consider solving the otherwise case. This needs to 1): dynamically recognize the breakpoints that {\em break} the sequential computation and cancel the related operations, and (2) select the results in appropriate ports since not all outputs are required. These are what we have additionally contributed to cope with. 

\subsubsection{\bf Accumulation Design for High-degree Vertex} 
There are also many high-degree vertices that over-fit the width of an accumulator. Invoking the accumulator multiple times can be considered a useful approach by dividing these edges into multiple parts and processing one of them at the same time, but this costs more overhead. 

First, iteratively reading the temporary vertex data and writing it back after merging with the accumulated result can lead to an extra synchronization. Second, the graph edges are sequentially stored with common data structure (e.g., {\em CSR/CSC} or {\em adjacency list}), which means that these edges are distributed to many continuous cachelines. When multiple vertices are simultaneously processed with a high-degree vertex, their edges may be located in non-adjacent cachelines, leading to  performance degradation.

We present a potential design with an efficient accumulation for solving these problems.

For the first problem, the update values of the same destination vertex come in sequence. It ensures that the results of multiple accumulations for the same high-degree vertex are also continuously generated. Therefore, the write back of the vertex data can be delayed before the accumulator sending a different vertex. 

For the second problem, the inefficiency mainly comes from fixed granularity for vertex scheduling. Without considering the differences in the vertex degree, it schedules fixed number of vertices and simultaneously accesses their edges in each cycle. Instead of accessing the edges based on the scheduled vertices, the viable method is to sequentially access all edges and dynamically schedule the vertices based on the accessed edges. 

\subsection{ Parallel Accumulator Architecture}

\begin{figure}
\centering
\includegraphics[width=3.2in, height=1.8in]{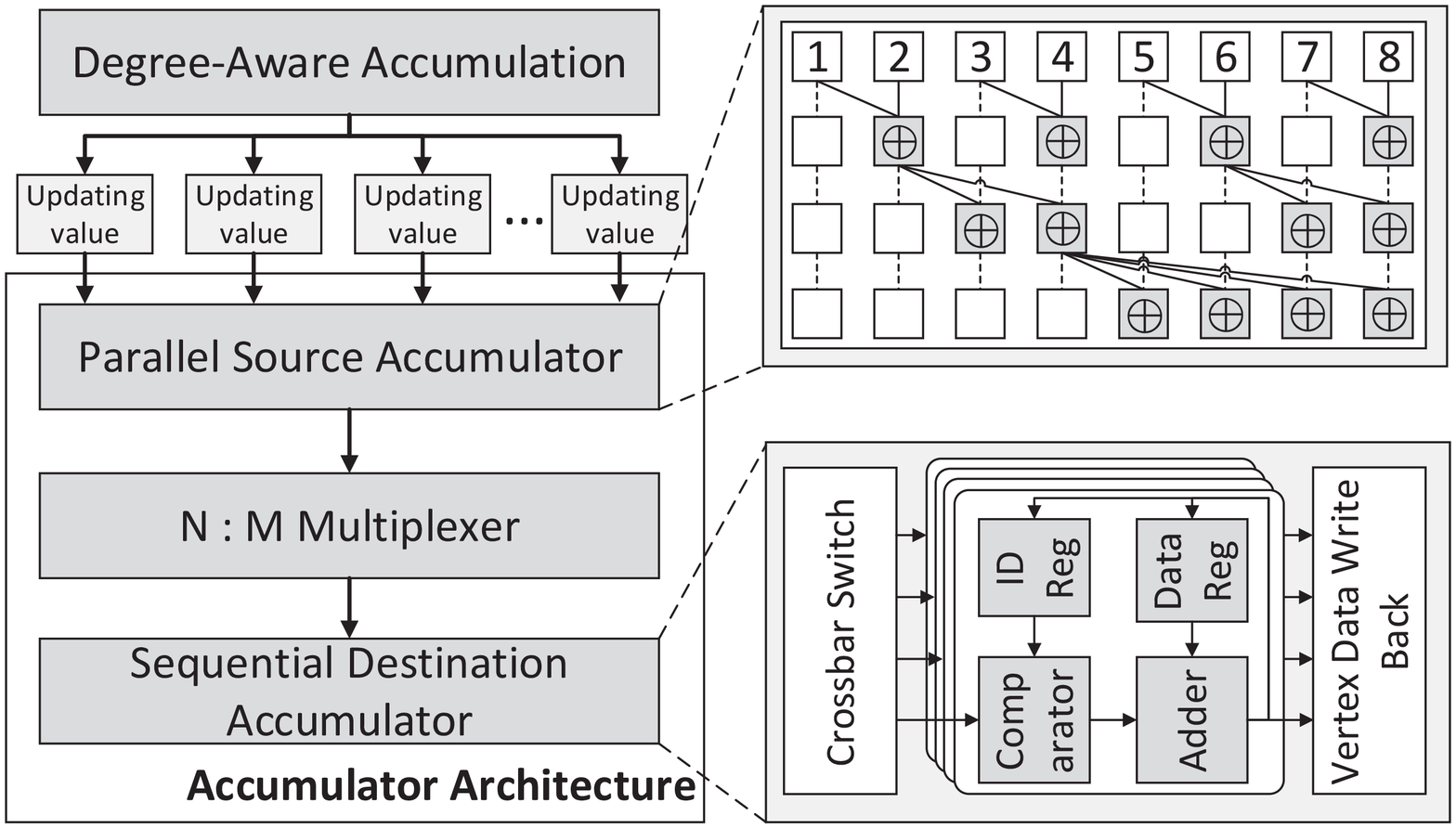}
\caption{Architecture of parallel accumulator}
\vspace{-1em}
\label{src_acc}
\end{figure}

Figure~\ref{src_acc} shows the overview of a parallel accumulator, consisting of four parts. The {\em source vertex accumulator} simultaneously accumulates update values of different destination vertices. The {\em multiplexer} is responsible for dynamically selecting accumulated data from appropriate ports of the source vertex accumulator. The {\em destination vertex accumulator} receives the selected data and fully accumulates each destination vertex. The {\em degree-aware accumulation} dynamically decides the number of vertices to be scheduled.

{\bf Source Vertex Accumulator: } The research of prefix-sum has been extensively studied since 1960s~\cite{sklansky1960conditional, kogge1973parallel, brent1982regular, ladner1980parallel}. In this work, we choose Ladner-Fischer Adder~\cite{ladner1980parallel} as the basis of our accumulator among a large number of previous efficient accumulators for three reasons as follow.

First, our main objective is to get the accumulated results in minimal latency, which filters the networks with depth larger than log($N$). Second, among all networks with minimal latency, it has relatively fewer adders, which means that we could add fewer extra resources for breakpoint recognition and result selection. Finally, although its fanouts are relatively larger than others, it does not increase the length of critical path since its delay and route time is much smaller comparing to that of on-chip memory access. 

Ladner-Fischer Adder opens a great opportunity for our graph-specific source vertex accumulator. In Ladner-Fischer Adder's original design, it establishes a fixed mapping relationship between the inputs and outputs, which leads to incorrect results when multiple vertices with mutable degrees are processed. As a result, we complement a breakpoint recognizing mechanism.  We add a new vector $V = (v_1, v_2, \ldots, v_N)$ where $v_i$ denotes the destination vertex that $a_i$ belongs to. With the vector $V$, the recognition conditions could be easily implemented by comparing the destination vertices of two inputs: 

\begin{align}
\label{new_dp}
f(i) = \left \{
\begin{aligned}
f(i - 1) + a_i, & \quad v_i = v_{i - 1} \\
a_i,            & \quad v_i \neq v_{i - 1}
\end{aligned}
\right.
\end{align}

We attach each update value with the ID of its destination vertex in our source vertex accumulator. To further reduce resource usage, we compress the destination vertex ID by only using its last log$m$ bits, where $m$ denotes the width of the accumulator. Based on Formula~(\ref{new_dp}), the adder nodes (refer to the gray nodes) are modified to compare the IDs of two inputs at first. If two IDs are the same, the behaviors of the adder nodes are the same as the original design which directly accumulates the input values. Otherwise, they will recognize the second destination vertex as breakpoint and send its update value to the output. 

{\bf Multiplexer: } Once the results are accumulated, the next is to dynamically select the accumulated results for each destination vertex from the output ports of source vertex accumulator. We use a $N \times M$ multiplexer to implement such a logic. Instead of directly comparing the destination vertex IDs, the multiplexer selects the data based on edge offsets to simplify the conditional logic. When the edges in pipeline stage P2 are accessed, each scheduled vertex is attached with its right edge offset, indicating the last edge connected to it. Based on this information, the multiplexer is thus able to naturally select the data for each scheduled destination vertex in the ports related to its last edge. For example, if the updated values $a_1, a_2, a_3$ belong to the vertex, the multiplexer would select the accumulated data from the third port of the source vertex accumulator.

\begin{figure}
\centering
\includegraphics[width=3.3in, height=1.2in]{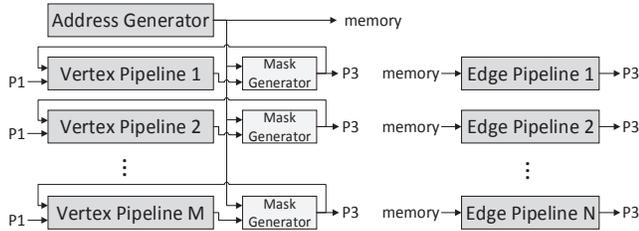}
\caption{Degree aware accumulation}
\vspace{-1.5em}
\label{fig_edge_parallel}
\end{figure}

{\bf Destination Vertex Accumulator: } In light of the sequential arrival of accumulated values, this opens an opportunity to avoid synchronization on the temporary vertex data by delaying the write back of the destination vertex data until the accumulated value of a different vertex is received. 

We design a destination vertex accumulator to merge different accumulated results of the same vertex. The accumulator holds the destination vertex ID and the accumulated value in private registers. In each cycle, if the IDs in the input and register are found to be the same, the accumulator would accumulate the vertex data in the input and register. Otherwise, the vertex data in the register will be written back and replaced by the input data. Furthermore, since the source vertex accumulator may simultaneously process multiple destination vertices, we replicate the destination vertex accumulators and use a crossbar switch to connect them with multiplexer. The crossbar switch routes the vertex data based on the destination vertex. That is, the last log($m$) bits in its ID are used for $m$ replications.

{\bf Degree Aware Accumulation: } Figure~\ref{fig_edge_parallel} shows the specific design of degree aware accumulation. The basic idea is to sequentially access all edges and dynamically schedule vertices based on the runtime information of their edge offsets (e.g., edge ID table in CSR/CSC~\cite{ham2016graphicionado} which denotes the location for the edges of each vertex). To make sure that multiple vertices could be accessed in each cycle, we replicate vertex units in stage P1 and P2. Furthermore, a special matching mechanism is implemented in the vertex units of stage P2 to dynamically decide the vertices to be scheduled.

More specifically, we use a specialized generator to automatically generate memory address to sequentially access all edges. In each cycle, every vertex unit stores received edge offsets, and compares generated memory address with the top data in its FIFO.  If the memory address is within the range of two edge offsets, the top vertex would be scheduled and sent to the next stage. Moreover, if the memory address is equal to the right edge offset, which means all edges of the vertex have been read, the top vertex in the FIFO would be removed. In this way, the number of scheduled vertex is ensured to be the same with that of vertex contained in requested cacheline. Furthermore, the edge units could be shared among all vertex units to improve resource utilization.

\section{Optimizations For Efficient Use}
In this section, we present several optimizations that are the key for using the proposed parallel accumulator efficiently.

\begin{figure}
\centering
\includegraphics[width=2.6in, height=1.5in]{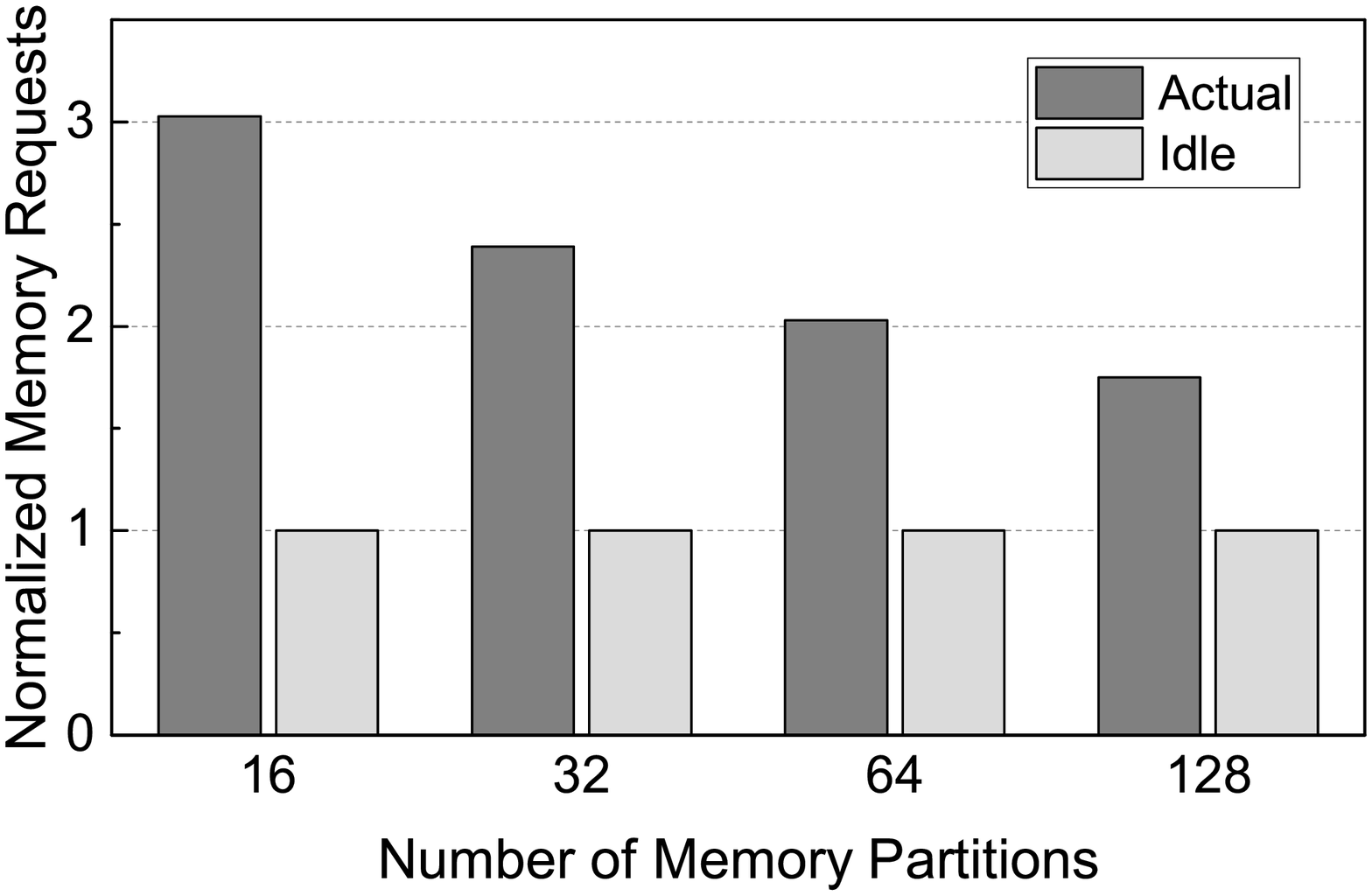}
\caption{Normalized performance for processing 16 random memory requests}
\vspace{-1.5em}
\label{fig_memory_partition_overhead}
\end{figure}

\subsection{Source Vertex Access Parallelization}
While the above accumulator can provide reasonable execution efficiency, the memory access is likely to be a potential performance bottleneck. In practice, the neighbors of every vertex are discontinuous, leading to significant randomness in vertex access. Consequently, the vertex data is typically stored in on-chip memory (e.g., BRAM in FPGA)~\cite{dai2017foregraph, nurvitadhi2014graphgen, ozdal2016energy} to improve memory performance. 

Despite that it could efficiently reduce the latency of vertex access, the throughput of on-chip memory is hard to keep with that of accumulator. For example, assuming that the accumulator runs at 250MHz with a DDR4-2400 memory. In each cycle, the accelerator would receive 16 32-bits edges and generate memory requests based on their source vertices, which means the on-chip memory need to simultaneously process 16 random read requests. Nevertheless, the standard RAM module could only process one read and write request in each cycle. Considering the limitation of capacity and frequency for on-chip memory in typical FPGA chips, memory partitioning~\cite{cong2011automatic, wang2013memory} is the most practical method to implement such multi-ported memory.

Typical memory partitioning mechanisms divide the memory into $n$ independent parts and shuffle the requests to achieve a maximal throughput of $n$. Nevertheless, due to the randomness in vertex access, we find a significant number of requests are shuffled to the same memory partition in each cycle, which means that the memory needs more than one cycle to process these requests. As shown in Figure~\ref{fig_memory_partition_overhead}, the unbalanced shuffling increases up to 70\% cycles, even if we partition the memory into 128 parts.

%


{\bf Optimizations: } Through analysing the graph data, we find that such inefficiency is caused by the unbalanced edge values: 1) the edge values are not evenly distributed when accessing in the cacheline-width granularity, 2) the edge values themselves are unbalanced when processing in the single-vertex granularity. 

Algorithm~\ref{alo_rearrange} represents the pseudocode of our mechanism for solving the first problem. The basic idea is to rearrange the edges of each vertex to ensure that the address values are relatively balanced in cacheline-width granularity before processing the graph. Assuming that the memory is partitioned to 16 dependent parts, we would also maintain 16 queues for each vertex to store the edges based on the connected vertex's ID. During rearranging, we would iteratively select edges from each queue in sequence for every vertex. The overhead of rearrangement is about O(|E|), which is the same as that of compressing algorithms commonly used in graph processing (e.g., CSR/CSC). With the mechanism, the address values could be evenly rearranged, thus improving the memory performance. 

\begin{figure}
\includegraphics[width=3in, height=1.6in]{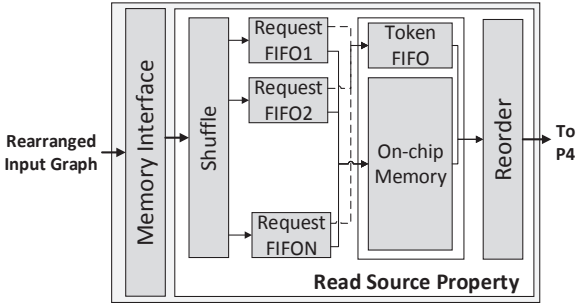}
\caption{Workflow of accessing source vertex data}
\vspace{-1.5em}
\label{fig_source_access}
\end{figure}

\begin{algorithm}
\caption{Pseudocode of the rearranging mechanism}
\label{alo_rearrange}
\small
\KwIn {Graph $G = (V, E)$, partition number $P$}
\KwOut {Rearranged edge list $NewEdge$}
\For {$v \in G$} {
    \For {$u \in \{k | (k, v) \in E \}$} {
        $Edge(v, u \ MOD \ P).push(u)$\;
    }
    $N(v) \leftarrow |\{k | (k, v) \in E \}|$\;
}
\For {$v \in G$} {
    $i \leftarrow 0$\;
    \While {$N(v) > 0$} {
        $NewEdge(v).push(Edge(v, i).pop())$\;
        $i \leftarrow (i + 1) \ MOD \ P$\;
    }
}
\end{algorithm}

For the second problem, we find that even though address values of single vertex are unbalanced, those of the whole graph are relatively balanced. Therefore, we try to change processing granularity to deal with such imbalance. More specifically, we allow the on-chip memory to process the requests in an unblocking (out-of-order) manner. Through unblocked process, the idle memory ports could be utilized by the latter requests, thus improving memory efficiency.


Figure~\ref{fig_source_access} shows the work flow of our mechanism. In each cycle, stage P3 receives $N$ edges from memory, and shuffles them to different request FIFOs based on their values. The FIFOs cache these edges and send the requests generated by the top ones to the on-chip memory. To avoid the unblocked requests breaking sequentiality of edge access and further leading to incorrect results, a reorder stage is involved after accessing the source vertex data. The reorder stage caches the accessed vertex data, reorders them to match the sequence of original requests, and sends reordered data to stage P4. To implement such reordering logic, each memory request would be attached with a token based on the last log$(m)$ of original edge memory address, where $m$ denoted the size of buffer in reorder stage. All accessed data with the same token would be stored in the same location in reorder stage. Once the top data finishes reordering, i.e., all data of the first request has been received, it would be sent to the next stage.

\subsection{Source-Based Graph Partition}
While storing vertex data in on-chip memory could avoid costly random access in main memory, it might require a large number of resources that exceed the capacity of the chip. Assuming the 4-byte width of vertex data and 8 M vertices, the on-chip memory is desired to be larger than 32 MB, which is unpractical for most of FPGAs. To enable process of large-scale graphs without losing the benefit of on-chip memory usage, we partition the graph into several parts and process a single part at a time.

To ensure that all vertex data needed to be processed in each graph parts could be held in on-chip memory, we use a source-based partition mechanism~\cite{gonzalez2012powergraph}. The partition mechanism works as follows. Firstly, the vertices of the input graph are divided into $K$ parts based on their vertex IDs. The value of $K$ depends on the number of vertex and the capacity of on-chip memory. For each part, the out-edges of each vertex are also included. After the input graph is partitioned, our accelerator sequentially processes each graph part in each iteration. Since every edge would be partitioned to the graph part which includes its destination vertex, no edges need to be processed twice. The graph partition does incur some extra memory overhead, since the same destination vertex data might be read and written more than once. More specific impacts would be discussed in Section 5.4.

\section{Evaluation}
This section evaluates the effectiveness and efficiency of our graph accelerator on a wide variety of graph algorithms with real-world graph datasets.

\subsection{Experimental Settings}
{\bf Evaluation Tools: } We implement our accelerator on Xilinx Virtex Ultrascale+ XCVU9P-FLGA2104 FPGA with -2L speed grade. The target FPGA chip provides 1.18 M LUTs, 2.36 M registers, and 9.49 MB on-chip BRAM resources. We verify the correctness and get the clock rate as well as resource utilization using Xilinx Vivado 2017.1. All these results have passed post-place-and-route simulations. Our target off-chip memory is Micron 4GB DDR4 SDRAM (MT40A256M16GE-083E). We use DRAMSim2~\cite{rosenfeld2011dramsim2} to simulate the cycle-accurate behavior of the off-chip access. The memory has a running frequency of 1.2 GHz and a peak bandwidth of 19.2 GB/s. 

\begin{table}[htbp]
\centering
\vspace{-0.5em}
\caption{Graph datasets} 
\vspace{-1.5em}
\label{Graph_datasets}
\tabcolsep=0.1cm
\begin{tabular}{|c|c|c|c|}
\multicolumn{1}{c}{}&\multicolumn{1}{c}{}&\multicolumn{1}{c}{}&\multicolumn{1}{c}{} \\ \hline
{\bf Names}   			& {\bf\# Vertices}	& {\bf\# Edges} 	& {\bf Description} \\ \hline
Slashdot 	& 0.08 M 			& 0.95 M			&	Link Graph \\ \hline
DBLP       	& 0.32 M   			& 1.05 M			&	Collaboration Graph \\ \hline
Youtube 	& 1.13 M			& 2.99 M			&	Social Network \\ \hline
Wiki		& 2.39 M			& 5.02 M			&	Website Graph \\ \hline
LiveJournal	& 4.85 M  			& 69.0 M			&	Follower Graph\\ \hline
Orkut    	& 3.07 M  			& 117 M				&	Social Network\\ \hline
\end{tabular}
\vspace{-0.5em}
\end{table}

{\bf Graph Algorithms: } We implement three well-known graph algorithms on our accelerator, covering both CAS-if and atomic-add operation types in Table~\ref{atomic_type}. 
\begin{itemize}[leftmargin=*]
    \item {\em Breadth First Search (BFS)} is a basic traversal algorithm utilized by many graph algorithms. It iteratively traverses the input graph and calculates the distance of shortest path from root to every vertex. 
    
    \item {\em PageRank (PR)} is an important graph algorithm used to rank web pages according to their importance. It updates every vertex based on the formula $Rank(v) = \varepsilon + \newline \sum_{u \in in-neighbor(v)} Rank(u) / |out-neighbor(u)|$ in each iteration, where $\varepsilon$ is a constant. 
    
    \item {\em Weakly Connected Components (WCC)} is an algorithm that checks the connectivity between two vertices in a graph. During the traverse, every vertex would receive the labels from all neighbors and update itself with the minimal one. 
\end{itemize}

{\bf Graph Datasets:} The graph datasets for the experiments are summarized in Table~\ref{Graph_datasets}. All these graphs are real graph data sets collected from SNAP~\cite{snapnets} and TAMU~\cite{DvisSparse}. In our implementation, each undirected edge is treated as two directed edges between source vertex and destination vertex by being processed twice. Therefore, the number of edges for undirected graphs ({\em DBLP}, {\em Youtube}, and {\em Orkut}) is considered double in our evaluation.

\begin{figure}
\centering
\includegraphics[width=2.8in, height=1.5in]{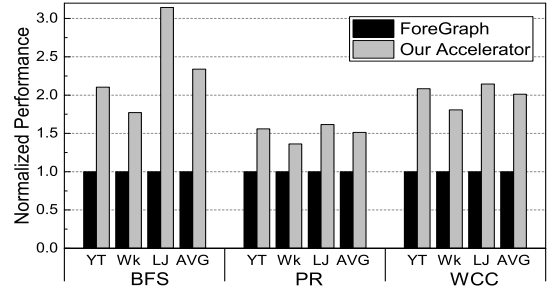}
\caption{Our accelerator normalized to the ForeGraph performance. YT denotes graph {\em Youtube}, Wk denotes graph {\em Wiki}, and LJ denotes graph {\em LiveJournal}. AVG presents the average speedup of all tested graphs}
\label{fig_performance_compare_foregraph}
\end{figure}

\subsection{Overall Performance}

{\bf Resource utilization: } 
Table~\ref{resource_utilization} shows the resource utilization and clock rate of the FPGA design with 8 vertex pipelines and 16 edge pipelines, which maximizes throughput given the peak DRAM bandwidth. First of all, because of the shared edge pipeline design described in Section 3.2, the number of resources required is reduced. Therefore, the logic resource (LUT and register) consumption of our accelerator is relatively low. Secondly, we implement the on-chip memory with BRAM resources to maintain vertex data. Similar to prior work~\cite{dai2017foregraph}, we use 1 byte integer to represent the depth value in BFS, single-precision floating point (4 bytes) in PR, and 4 bytes integer in WCC. In this way, the maximal memory requirement is 1 $\times$ 4.85 $=$ 4.85 MB for 1 byte data and 4 $\times$ 4.85 $=$ 19.4 MB for 4 bytes data. Therefore, we hold all vertex data when running BFS and about 1.7 M vertex data for other algorithms, which consumes 57.9\% and 69.9\% of available BRAM resources, respectively. The UltraRAM resources are not used in our implementation.

\begin{table}[htbp]
\centering
\caption{Resource utilization and clock rate} 
\vspace{-1.5em}
\label{resource_utilization}
\begin{tabular}{|c|c|c|c|}
\multicolumn{1}{c}{}&\multicolumn{1}{c}{}&\multicolumn{1}{c}{}&\multicolumn{1}{c}{} \\ \hline
{\bf }   				& {\bf BFS}		& {\bf PR} 		& {\bf WCC} \\ \hline
LUT 					& 7.39\% 		& 10.1\%		& 8.26\% \\ \hline
registers       		& 2.53\%   		& 4.47\%		& 3.02\% \\ \hline
BRAM 					& 57.9\%		& 69.9\%		& 69.9\% \\ \hline
Maximal clock rate		& 256 MHz  		& 211 MHz		& 251 MHz \\ \hline
Simulation clock rate   & 250 MHz  		& 200 MHz		& 250 MHz \\ \hline
\end{tabular}
\vspace{-0.5em}
\end{table}


{\bf Throughput:} Figure~\ref{fig_performance_compare_foregraph} shows the normalized performance comparing to ForeGraph, which is one of the fastest graph processing accelerator implemented on FPGA, with respect to throughput. By throughput, we refer to the number of {\em traversed edges per second} (TEPS)~\cite{Graph500}, which is a performance metric frequently used in graph processing. As described above, ForeGraph is a representative accelerator that sequentially processes different edges of the same destination vertex to ensure atomicity. 

Since ForeGraph has not been open-sourced, we execute the same graph algorithms (BFS, PR, and WCC) and datasets ({\em youtube}, {\em wiki-talk} and {\em LiveJournal}) used by its evaluation on our accelerator, and compare the results with the performance reported in its work (just as previous work has also done~\cite{dai2017foregraph, zhou2016high}). When running PR and WCC on {\em Wiki}, the BRAM resources available in the FPGA chip used in ForeGraph is large enough to (up to 16.6 MB) hold all vertex data on-chip, which is unreliable for that of our FPGA chip (9.49 MB). Therefore, we compress the vertex data to 2 bytes when running PR and WCC on {\em Wiki} for fair comparison.

As shown in Figure~\ref{fig_performance_compare_foregraph}, our accelerator achieves 1.36x $\sim$ 3.14x speedup compared to the ForeGraph. As analysed in Section 2.2, the speedup mainly comes from the reduced synchronization overheads by simultaneously processing atomic operations. Moreover, our accelerator could achieve better load-balance using degree-aware accumulation by dynamically deciding the number of vertices scheduled. 

For the results of different algorithms, we find that the speedup of PR is smaller. This is because of the lower clock rate caused by complex floating units. Since the number of edge pipelines is fixed in our implementation, the clock rate directly influences the overall performance. Moreover, the floating point units significantly increase the length of pipelines, thus would need more cycles when recovering from pipeline stalls. Therefore, the algorithms that use integer values could achieve slightly higher performance.
 
\subsection{Sensitivity Study}
\begin{figure}
\centering
\subfigure[Different graphs]{
\begin{minipage}[b]{0.46\linewidth}
\includegraphics[width=1.6in, height=1.2in]{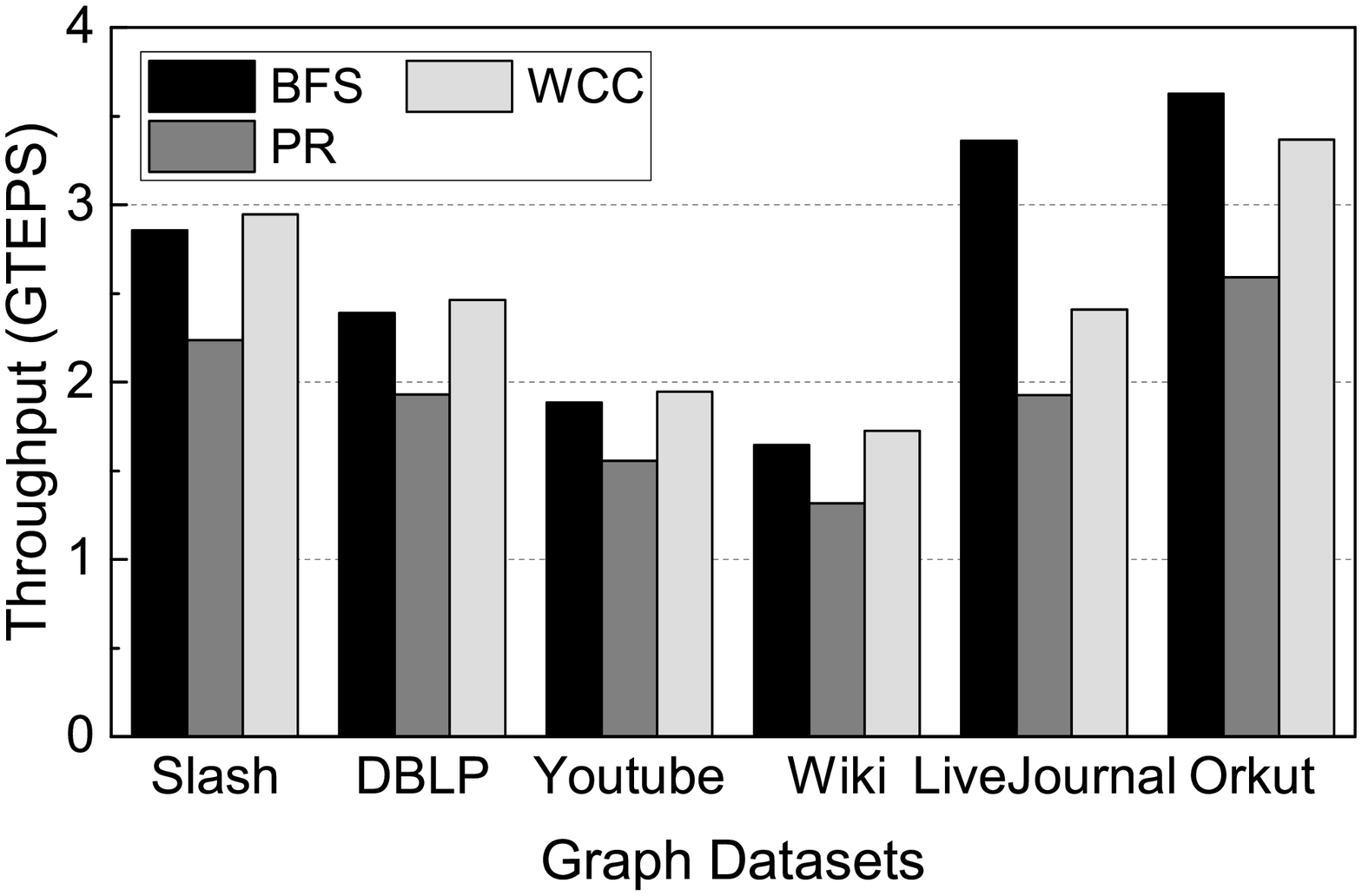}
\label{fig_throughput}
\vspace{-1.5em}
\end{minipage}
}
\subfigure[Different average degrees]{
\begin{minipage}[b]{0.45\linewidth}
\includegraphics[width=1.6in, height=1.2in]{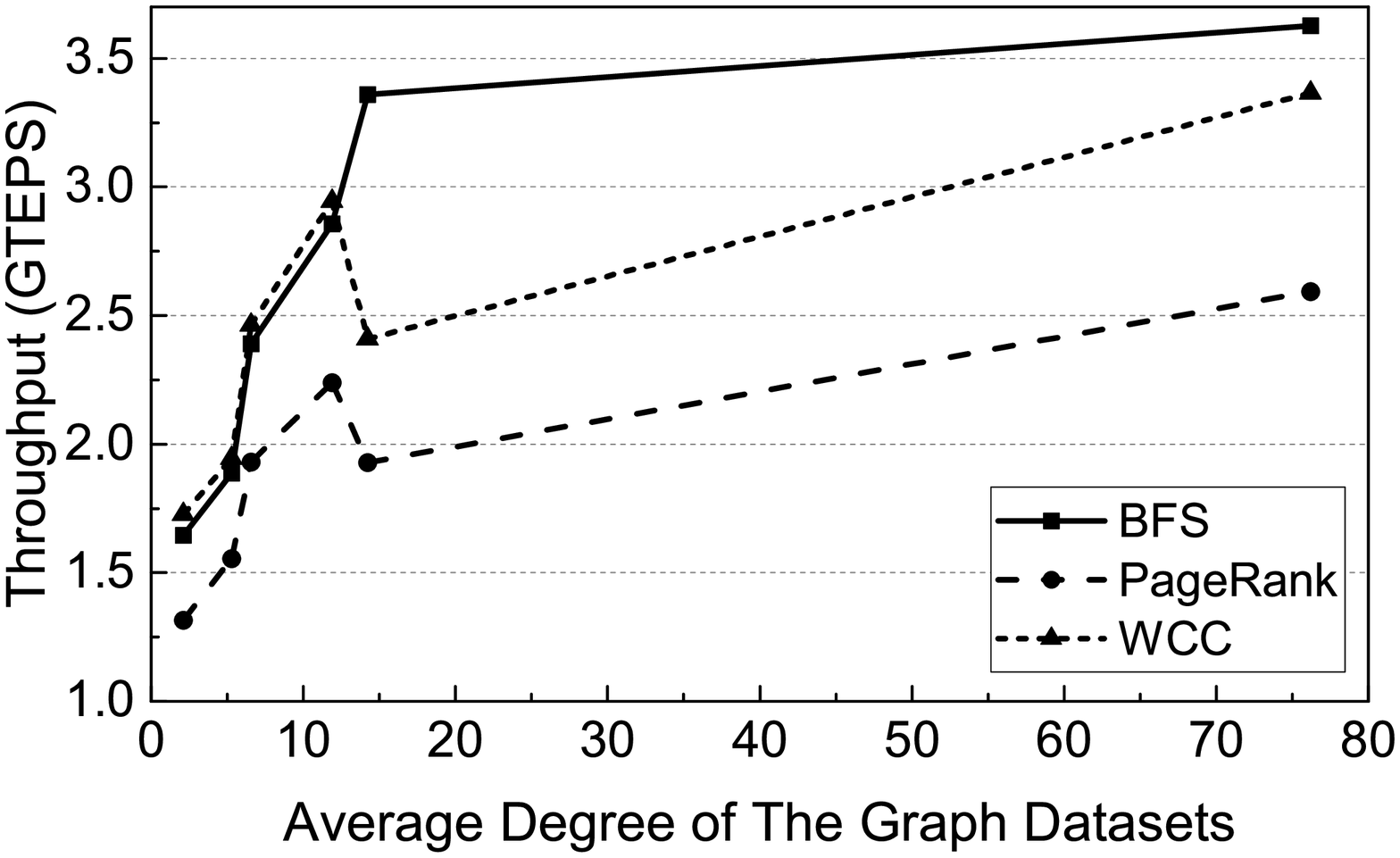}
\label{fig_averge_throughput}
\vspace{-1.5em}
\end{minipage}
}
\vspace{-1em}
\caption{Sensitive study on throughput with different graphs and average degrees} 
\vspace{-2em}
\end{figure}

To get a more comprehensive performance result, we execute all graphs described in Table~\ref{Graph_datasets} on our accelerator. The structures of these graphs significantly differ from each other (e.g., number of vertices and edges, average degree), thus providing an in-depth overview on the performance. As shown in Figure~\ref{fig_throughput}, our accelerator achieves 1.4 GTEPS $\sim$ 3.5 GTEPS over all graph algorithms and datasets.


Among all graph datasets, {\em Wiki}'s throughput is particularly low when executing on our accelerator. This is because {\em Wiki} is extremely sparse and makes the accelerator exhibits unbalance between the vertex and edge pipelines. With low average, the edges accessed from {\em Wiki} in each cycle prefer to belong to multiple vertices (more than 8). Therefore, the vertex pipelines might need more than one cycle to process these edges, leading to lower performance. 

As shown in Figure~\ref{fig_averge_throughput}, the performance is almost linearly increased when the average degree is less than 16. This is because that the percentage of low-degree vertex ($\le 2$) decreases. Moreover, the performance improves slightly when increasing the average degree from 16 to 76. This is because that the memory bandwidth becomes the potential bottleneck in these cases, since it could only send a cacheline-width edges in each cycle. In summary, the performance improves as the average degree increases before reaching the limitation of maximal memory bandwidth. 

Lastly, we find obvious performance degradation for PR and WCC when average degree is about 14 ({\em LiveJournal}). Moreover, the performance of PR and WCC is significantly lower than that of BFS when average degree is larger than 14 ({\em LiveJournal} and {\em Orkut}). This is because that the vertices data is too large to be all held in on-chip memory in these cases. Therefore, the graph partition mechanism is used when executing PR and WCC on these graphs, which involves in more vertex access. More detailed analysis of degree distribution and graph partition is presented in Section 5.4.

\subsection{Benefit Breakdown}
\begin{figure}
\centering
\includegraphics[width=2.6in, height=1.5in]{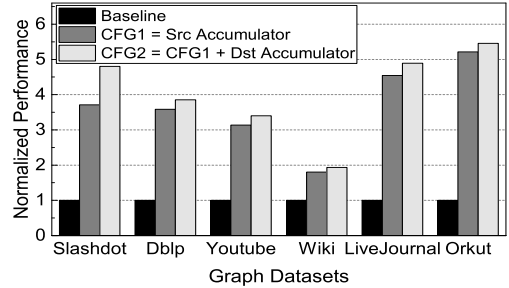}
\caption{Benefit of parallel accumulation}
\vspace{-1.5em}
\label{performance_accumulator}
\end{figure}

We next break down the respective benefits of our different graph accelerator designs as follow:

{\bf Benefits from Parallel Accumulation: } Figure~\ref{performance_accumulator} presents the normalized performance results. 
The baseline represents the basic design without any optimizations described in Section 3 and 4. It sequentially processes each edge, and accumulates its values to the final result in each cycle. CFG 1 represents source vertex accumulation. CFG 2 further uses destination vertex accumulation based on CFG1. 

It is shown that CFG1 achieves 1.9x$\sim$ 5.2x speedup compared to the baseline. Note that {\em Wiki} is lowest performance among all graph workloads. This is because that the number of vertex pipelines to set to one, leading to the fact that only one vertex can be scheduled in each cycle for CFG 1. Therefore, the number of edges sent to the accumulator in each cycle is directly depended on the average degree. In a word, the graphs with higher degree could experience higher speedup when using source vertex accumulator.

For CFG 2, destination vertex accumulator achieves about 1.3x speedup in most of graphs, except for {\em Slashdot} (2.0x speedup). This is because that {\em Slashdot} has self-loops, which means that some edges connect a vertex to itself. When processing these self-loops, the memory requests of source and destination vertex would be assigned to the same on-chip memory partition, leading to increased memory cycles. With the source vertex accumulator, the request of destination vertex could be avoid, thus improving the overall performance.

\begin{figure}
\centering
\subfigure[Performance]{
\begin{minipage}{0.42\linewidth}
\centering
\includegraphics[width=1.6in, height=1.2in]{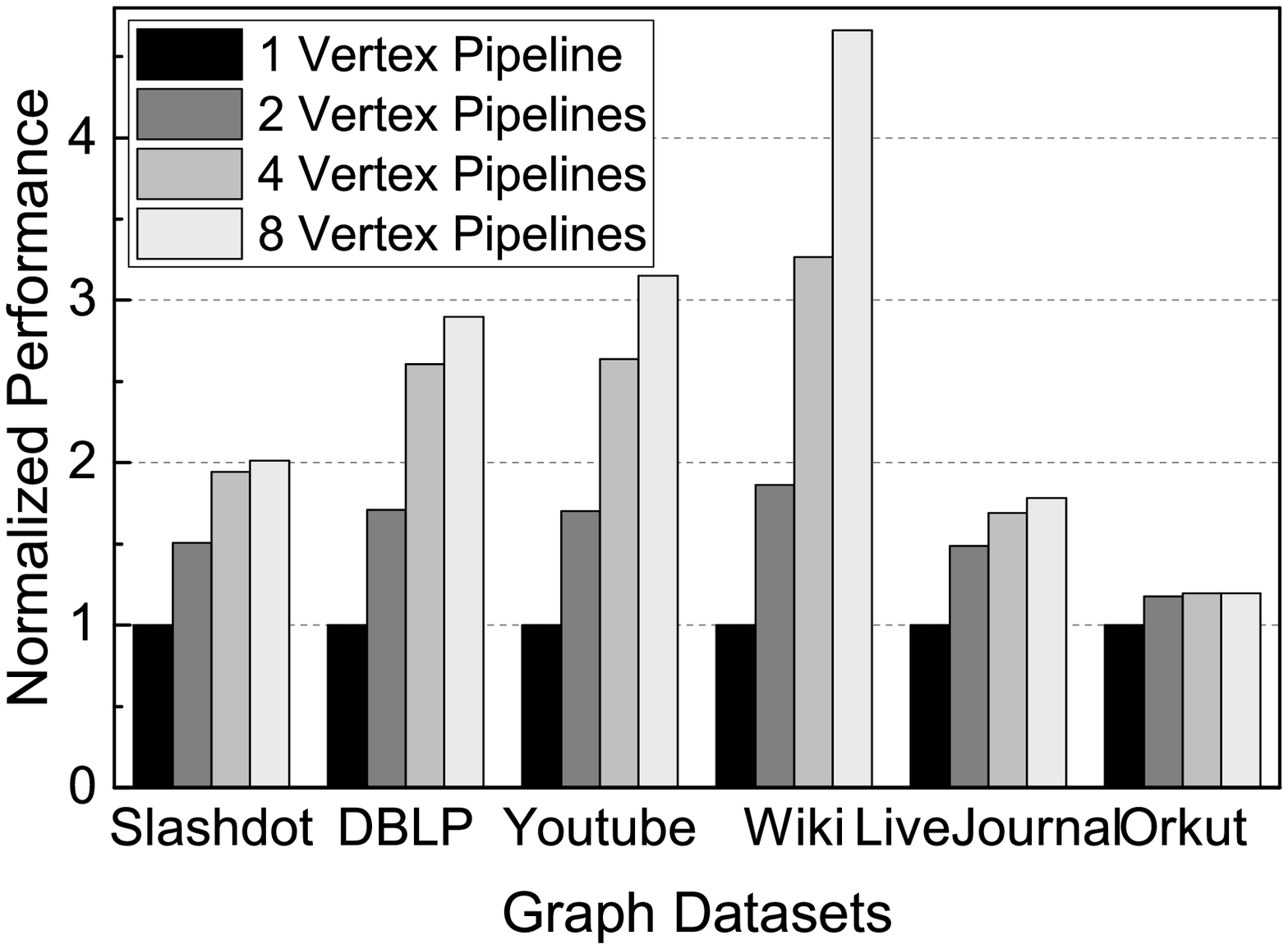}
\label{fig_performance_vertex_parallel}
\end{minipage}
}
\hspace{0.1in}
\subfigure[Percentage of covered edges]{
\begin{minipage}{0.42\linewidth}
\centering
\includegraphics[width=1.6in, height=1.2in]{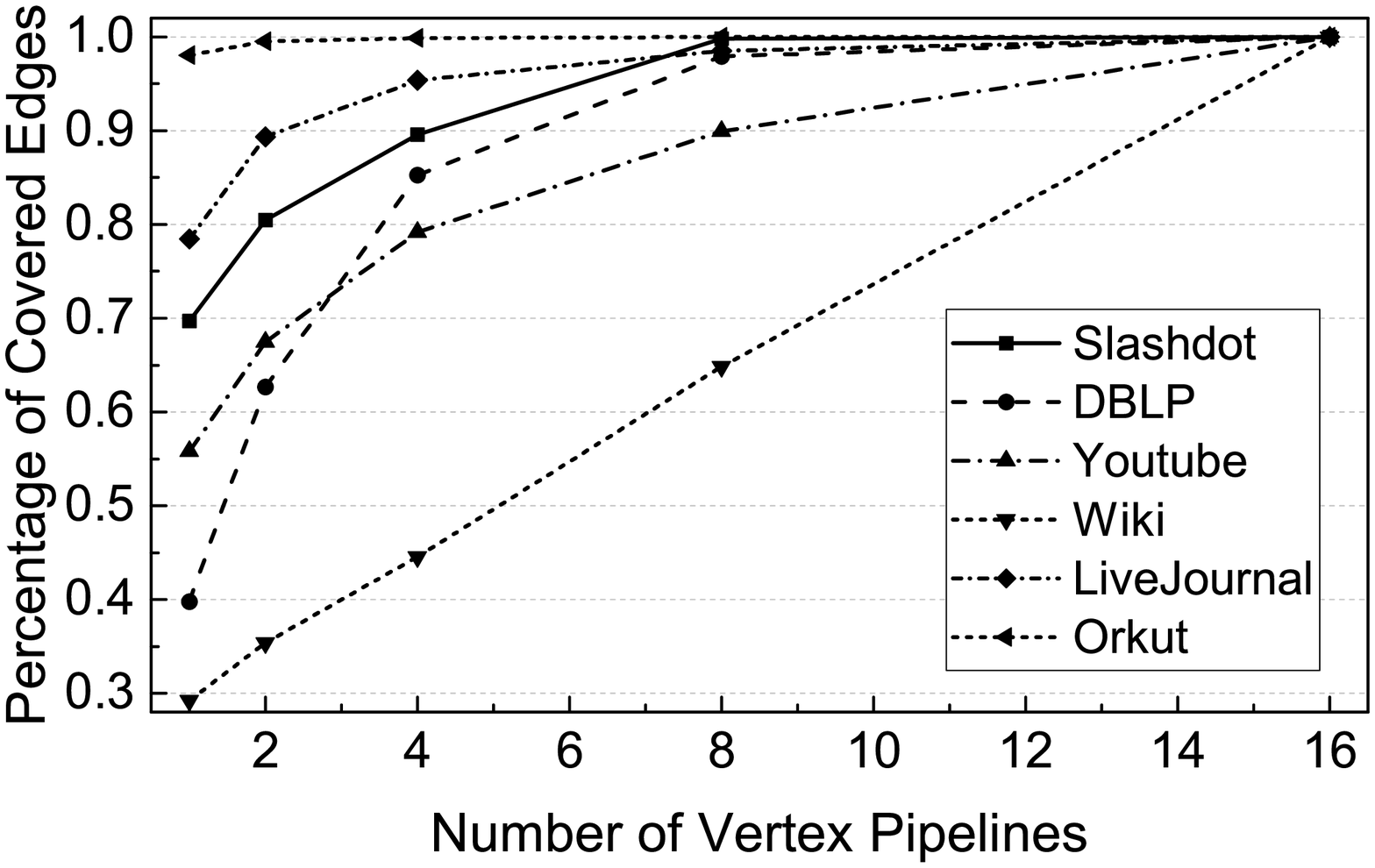}
\label{degree_distribution}
\end{minipage}
}
\caption{Benefit of degree-aware accumulation}
\vspace{-1.5em}
\end{figure}

{\bf Benefits from Degree-aware Accumulation: }Secondly, we explore the impact of degree aware accumulation on above accumulators. Figure~\ref{fig_performance_vertex_parallel} presents the results which assume that on-chip memory could process any 16 memory requests in each cycle. For the performance, we analyse the speedup brought by different number of vertex pipelines, which denotes the maximal parallelism of the accumulation\footnote{When the number of vertex pipelines is set to $N$, the mechanism dynamically schedules $1 \sim N$ vertices based on the degree.}. 


We make the observation that the performance improves sub-linearly as the number of vertex pipelines increases. This is because of the power-law degree distribution of graphs. Assuming that the number of vertex pipelines is $N$, our degree aware mechanism could cover the vertices with degree $\ge 16 / N$ with 16 edge pipelines. As depicted in Figure~\ref{degree_distribution}, the percentage of the covered edges for most graphs increases sub-linearly because that high-degree vertices have most of the edges. While for {\em Wiki}, the skewness of its degree distribution is low, thus leading to an almost linear increment.


\begin{figure}
\centering
\includegraphics[width=2.6in, height=1.5in]{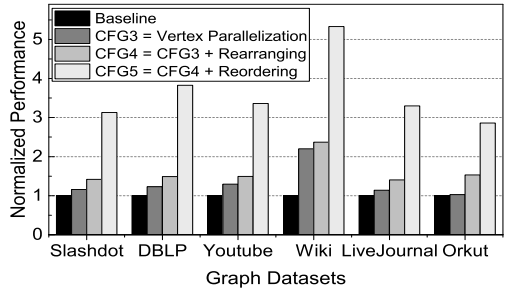}
\caption{Effect of different optimizations in memory subsystem discussed in Section 4}
\vspace{-1.6em}
\label{fig_performance_memory}
\end{figure}

{\bf Benefits from Vertex Access Parallelization: } Figure~\ref{fig_performance_memory} explores the impact of different optimization for parallel accumulations, without ignoring the influence of the on-chip memory's throughput. The left most bar in Figure~\ref{fig_performance_memory} represents the baseline case where only parallel accumulation is applied. CFG 3 represents that degree aware accumulation is involved in with 8 vertex pipelines based on CFG2. CFG 4 shows the effects of rearranging mechanism and CFG 5 shows the effects of reordering discussed in Section 4.1. 

The first observation is that the speedup of degree aware accumulation is decreased to about 1.3x when considering the influence of on-chip memory's throughput. Without any optimizations, there would be a significant amount of increased memory requests caused by the unbalanced edge values, thus decreasing the impact of degree aware accumulation. Another observation is that our rearranging mechanism could achieve 1.5x speedup and reordering mechanism could achieve another 1.5x $\sim$ 2.8x speedup. With these mechanisms, the increased memory requests could be reduced to $\le 10\%$, which significantly improves the memory efficiency.

\begin{figure}
\centering
\includegraphics[width=2.6in, height=1.5in]{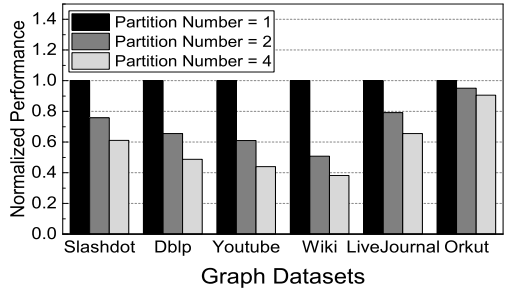}
\caption{Effect of graph partition mechanism}
\vspace{-1.5em}
\label{fig_performance_partition}
\end{figure}

{\bf Benefits from Graph Partition: } Figure~\ref{fig_performance_partition} explores the impact of graph partition described in Section 4.2. The leftmost bar represents the case where the on-chip memory size is enough to hold all vertex data, denoted as partition number = 1. The other bars represent cases where on-chip memory size is only enough to hold $1 / N$ of the total vertex data where $N$ represents the number of partitions.  

In general, partitioning the graphs into 4 parts would result in around 40\% performance degradation. Among all workloads, the {\em Wiki} experiences the largest performance degradation which reaches about 61\%. This is because that we would traverse all vertex in each sub-iteration when processing each graph partition. As the average degree decreases, the increased vertex access overheads would account for a significant percentage of total overheads. Therefore, the performance of graphs with lower average degree would be more sensitive to the partition number.

\section{Related Work}
A wealth of recent studies~\cite{guo2014well, beamer2015locality, guo2015empirical} indicate that even with extensive optimizations, graph processing still subjects to the underlying limitation of general-purpose processors. A vast body of research efforts have been therefore put into making the graph-specific architectural innovations to improve the execution efficiency. Graphicionado~\cite{ham2016graphicionado} proposes a pipelined graph accelerator which efficiently utilizes large on-chip scratchpad memory. GraphGen and Graphops~\cite{nurvitadhi2014graphgen, oguntebi2016graphops} propose FPGA-based frameworks which automatically compile graph algorithms to specialized graph processors. Compared with these prior researches with strict atomic protection, we argue that the heavy reliance on atomic operations leads to significant performance degradation and propose a novel accelerator to reduce atomic overhead.  



There are also a large number of attempts that aim at reducing the number or the execution time of atomic operations for graph processing. ForeGraph~\cite{dai2017foregraph} partitions the input graph in a grid-manner~\cite{zhu2015gridgraph} to avoid simultaneously scheduling edges with the same vertex. ~\cite{ozdal2016energy} proposes a specialized synchronizing mechanism to avoid scheduling conflicting edges. Shijie et al~\cite{zhou2016high} use a combing network to avoid the same vertex being simultaneously scheduled through filtering the unnecessary edges before processing. In general, their basic idea is to avoid scheduling the edges with conflict vertices through preprocessing. Speculative Lock Elision~\cite{rajwar2001speculative} speculatively remove the lock operations and enable highly
concurrent execution.
As a comparison, we focus on the performance impact between multiple atomic operations, instead of the performance of atomic operation itself. We find that these atomic operations could be parallelized according to distinct characteristics in vertex updates of graph processing. We thus propose an efficient graph-specific accumulator to exploit the potential benefits of this insight. 


Many other efforts also have been put into improving the execution time of atomic operations. 
Tesseract~\cite{ahn2016scalable} offloads all graph operations to memory-based accelerator to ensure atomicity without requiring software synchronization primitives. There are also some researches~\cite{ahn2015pim, nai2017graphpim} enables offloading operations at instruction-level. They statically or dynamically detect the atomic instructions during processing and directly map them into PIM region~ with minor extension to the host processors. Compared to these PIM-enabled graph architecture, our accelerator can achieve efficient management on shared data conflicts without introducing special memory components. Moreover, our parallel data conflict management can be also integrated into PIM-enabled graph accelerators and help to reduce the memory requests.

\section{Conclusion}
In this paper, we present a pipelined graph processing accelerator to enable massive parallelism of vertex updates. Our accelerator provides a parallel accumulator to simultaneously schedule and process multiple destination vertices without losing edge-level parallelism. Moreover, the accumulator is designed to be degree-aware and can adaptively adjust the vertex parallelism to different kinds of graphs. We also present vertex access parallelization and source-based graph partition for better supporting the efficient use of graph accelerator.  Our evaluation on a variety of graph algorithms shows that our accelerator can achieve the throughput by 2.36 GTEPS on average, and up to 3.14x speedup compared to the stat-of-the-art FPGA-based graph accelerator ForeGraph with its single-chip version. 


{
\bibliographystyle{IEEEtran}
\bibliography{sample-bibliography}
}
\end{document}